\documentclass[aps,prc,floatfix]{revtex4}
\usepackage{amsmath}
\usepackage{graphicx}
\usepackage{aas_macros}
\usepackage[dvips]{color}
\usepackage[normalem]{ulem}  % \sout{old text} for strikeout
\newcommand{\sla}{\! \not \!}
\def\){\right)}
\def\({\left(}
\def\]{\right]}
\def\[{\left[}

\newcommand{\be}{\begin{equation}}
\newcommand{\ee}{\end{equation}}

\newcommand{\lsim}{\, \, \raisebox{-0.8ex}{$\stackrel{\textstyle <}{\sim}$ }}

\newcommand{\roughly}[1]%
{\mathrel{\raise.4ex\hbox{$#1$\kern-.75em\lower1ex\hbox{$\sim$}}}}

\newcommand\beq{\begin{eqnarray}}
\newcommand\eeq{\end{eqnarray}}

\def\Dsl{\,\raise.15ex \hbox{/}\mkern-12.8mu D}

\def\fm3{fm$^{-3}$}

\begin{document}
%\begin{frontmatter}
%
\preprint{\vbox{\hbox{LA-UR-07-XXXX}}}

\title{Superfluid Response and the Neutrino Emissivity of Neutron Matter}
\author{Andrew W. Steiner}
\affiliation{Joint Institute for Nuclear Astrophysics, National
Superconducting Cyclotron Laboratory and,  \\ Department of Physics
and Astronomy, Michigan State University, East Lansing, MI 48824}

\author{Sanjay Reddy}
\affiliation{Theoretical Division, Los Alamos National Laboratory, Los
Alamos, NM 87545}
\begin{abstract}
{We calculate the neutrino emissivity of superfluid neutron matter in the inner crust of neutron stars. We find that neutrino emission due to fluctuations resulting from the formation of Cooper pairs at finite temperature is highly suppressed in non-relativistic systems. This suppression of the pair breaking emissivity in a simplified model of neutron matter with interactions  that conserve spin is of the order of $v_F^4$ for density fluctuations and $v_F^2$ for spin fluctuations, where $v_F$ is the Fermi velocity of neutrons. The larger suppression of density fluctuations arises because the dipole moment of the density distribution of a single component 
system does not vary in time.  For this reason, we find that the axial current response (spin fluctuations) dominates.  
In more realistic models of neutron matter which include tensor interactions where the neutron spin is not conserved,  neutrino radiation from bremsstrahlung reactions occurs at order $v_F^0$. Consequently, even with the suppression factors due to superfluidity, this rate dominates near $T_C$. Present calculations of the pair-breaking emissivity are incomplete because they neglect the tensor component of the nucleon-nucleon interaction.}
\end{abstract}
\maketitle
%21.65.+f:Nuclear Matter, exotic molecules etc.
%26.60.+c: Nuclear matter aspects of neutron stars
%74.20.Fg: BCS theory and its development
%\section{Introduction}
%==============================================
\section{Introduction}

The long term ($10^5-10^6$ yrs) cooling of isolated neutron stars and
the thermal evolution of accreting neutron stars in binary systems are
sensitive to the weak interaction rates at high density. A recent review of the theory, modeling and observational constraints on neutron star thermal evolution can be found in Ref. \cite{YakovlevPethick:2004}. Here, we focus on a specific neutrino process which is expected to be relevant in superfluid neutron matter.

In superfluids, Cooper pairs break and recombine constantly at finite
temperature. Such processes can dominate the density and spin-density
fluctuations. In pioneering work, several decades earlier, Flowers,
Ruderman and Sutherland recognized that these fluctuations can couple
to neutrinos through the weak neutral current. They showed that when
the temperature was less than but comparable to the critical
temperature for superfluidity, neutrino emission due to the Cooper pair recombination processes was important \cite{FRS:1976}. Subsequently, this process which is now commonly referred to as the pair-breaking and formation (PBF) process and was recomputed in Refs. \cite{VoskresenskySenatorov:1987,Yakovlev:1999}.
Its role in the thermal evolution of isolated neutron stars was studied and shown to important \cite{Page:1998,Yakovlev:1998,Page04}.

Another context in which the PBF process plays a role is in accreting neutron stars which exhibit x-ray bursts and superbursts.  Current models for superbursts indicate that they arise due to unstable burning of carbon in the ocean of accreting neutron stars
\cite{Cumming:2001}. Agreement between theoretical models
(for the light-curves and recurrence times) and observation relies on
the assumption that carbon is ignited at a column-depth of about
$10^{12}$ g/cm$^2$. However, to ignite carbon at this depth the 
temperature there should be  $\simeq 5 \times 10^8$ K
\cite{Brown:2004,Cumming:2006}. Consequently the ignition condition and hence the recurrence time for superbursts are sensitive to the temperature profile of the crust in accreting systems. The temperature profile in turn depends on the balance between heating in the crust due to electron captures and pycnonuclear reactions \cite{Haensel:1990} and neutrino cooling \cite{Brown:2004}.

The inner crust of the neutron star is expected to contain a neutron
superfluid. When the matter density exceeds the neutron drip density ($4\times 10^{12}$ g/cm$^3$) a relatively low-density neutron liquid
coexists with a lattice of nuclei \cite{Pethick:1995}. Here, the
attractive s-wave interaction between neutrons induces superfluidity.
The pairing-gap rises from zero at neutron drip to a maximum
of about 1 MeV when the neutron fermi momentum $k_F \sim 200$ MeV and  then decreases to zero in the vicinity of the crust-core interface. For temperatures of relevance to the accreting neutron stars, it was found that the PBF process in the neutron superfluid resulted in rapid neutrino losses and cooled the crust to temperatures below those required for carbon ignition at the favored depth \cite{Cumming:2006}. Subsequently, additional heating processes in the outer crust due to electron captures on nuclei were shown to be relevant but were unable to produce the necessary heating in models which included PBF process in the crust \cite{Gupta:2007}.

The preceding discussion motivates a detailed investigation of the
neutrino emissivity arising due to the PBF process in neutron matter in the inner crust. Recently, this was recalculated by Leinson and Perez who find that earlier calculations violated vector-current conservation \cite{LeinsonPerez:2006a,LeinsonPerez:2006b}.
An improved treatment that satisfies current conservation yielded a result that was suppressed by the factor $v_F^4$ where $v_F=k_F/M$ is the neutron fermi velocity, $k_F$ is the neutron fermi momentum and $M$ is neutron mass. In the neutron star crust, where $v_F\sim 0.1$ this suppression is significant. Subsequently, Sedrakian, Muther and Schuck also calculated the PBF rate using an improved treatment
based on Landau Fermi liquid theory and found that it was suppressed by the factor $\sim T/M$ where $T$ is the temperature \cite{Sedrakian:2006}. This suppression is parametrically different from that obtained in Ref.~\cite{LeinsonPerez:2006a}. The PBF rate was also recently examined in Ref.~\cite{Kolomeitsev08} where the authors also accounted for Fermi liquid effects in superfluid neutron matter using the Larkin-Migdal-Leggett formalism~\cite{Larkin63,Leggett65}. They find that  the vector current response is suppressed by the factor $v_F^4$ in agreement with the finding of Leinson and Perez. 

In this article, we reexamine the nature of density and spin-density fluctuations in superfluid neutron matter using a simplified nuclear Hamiltonian. Our main findings are 
\begin{enumerate} 
\item{The spectrum of density fluctuations is suppressed at order $v_F^4$ in pure neutron matter in agreement with the findings of Leinson and Perez. } 
\item{The $v_F^4$ suppression is not generic and is {\it not} a consequence of vector current conservation. It is specific to simple one component systems where all particles have the same (weak) charge to mass ratio. In multicomponent systems such as the neutron star crust where neutrons coexist and interact with nuclei, the density fluctuations of the neutron superfluid occur at order $v_F^2$. }
\item{ For the case of simple nuclear Hamiltonian with only central interactions that conserve spin,  spin-density fluctuations occur at order $v_F^2$ and the these fluctuation dominate the neutrino emissivity. }
\item{ For the case of realistic nuclear interactions which contain a strong tensor component, spin is {\it not} conserved,  and spin fluctuations arise at order $v_F^0$. This feature is well known in the context of neutrino emission from neutron-neutron bremsstrahlung. We find that the bremsstrahlung rate continues to be the dominant neutrino emission mechanism even in the superfluid state for $T \ge ~T_C/5$. }     
\end{enumerate}

The article is organized as follows. We begin by discussing  the relation between the neutrino emissivity and the density and spin-density response functions. This is followed by a detailed investigation of the density-density response function and the role of vertex corrections in the superfluid state. Here we show that the vertex corrections required by conservation laws strongly suppress the response relative to the predictions of mean field theory as suggested in earlier work. Subsequently we discuss the spin-density response function and show that it dominates over the density response. Finally, we will discuss the various contributions to the neutrino emissivity and conclude that neutron-neutron bremsstrahlung rate is typically larger PBF process even in the superfluid phase. We conclude with a critical discussion of our study here and related earlier work. We recognize that all calculations of the neutrino rates from PBF are missing a key aspect of the nuclear force - namely the tensor interaction.     
   
\section{Neutrino Emissivity and Response Functions} The neutrino
emissivity is defined as the rate of energy loss per unit volume and is given by
\begin{eqnarray}
\dot{\epsilon}_{\nu\bar{\nu}}&=& - \frac{G_F^2}{4}
\int\!\frac{d^3q_1}{(2\pi)^32 \omega_1}
\int\!\frac{d^3 q_2}{(2\pi)^3 2\omega_2} ~ \int\! d^4 \vec{k} \quad \delta^4(\vec{q}_1 + \vec{q}_2 -  \vec{k}) 
\quad \frac{\omega}{\exp{\left(\beta \omega\right)}-1} ~
 L^{\alpha\beta}(q_1,q_2)\quad \Im m[\Pi^R_{\alpha\beta}(k)],
 \label{eqn_emis1}
 \end{eqnarray}
where $G_F$ is the Fermi weak coupling constant, $k =
(\omega,\vec{k})$, $q_{i=1,2}$ are the on-mass-shell four-momenta of
neutrinos, $L^{\alpha\beta}(q_1,q_2) = {\rm Tr}\left[\gamma^{\mu}(1 -
\gamma^5)\sla{q_1} \gamma^{\nu}(1-\gamma^5)\sla{q_2}\right]$ and
$\Pi^R_{\alpha\beta}(q)$ is the retarded polarization
tensor~\cite{Sedrakian:2006}. Using Lenard's identity
\cite{Lenard:1953}, we can simplify Eq.~\ref{eqn_emis1} to obtain
\begin{eqnarray}
\dot{\epsilon}_{\nu\bar{\nu}}&=& \frac{G_F^2}{192~\pi^5} 
\int\! d^3 \vec{k}\int_0^{\infty}\! d \omega
~\Theta[\omega^2-|\vec{k}|^2] ~\left( k^{\alpha}k^{\beta} - k^2 g^{\alpha \beta}\right)
\quad  \frac{\omega}{\exp{\left(\beta \omega\right)} -1} 
~R_{\alpha\beta}(-\omega,|\vec{k}|)\, 
\end{eqnarray} 
where the superfluid response function
$R_{\alpha\beta}(\omega,|\vec{k}|)$ in general contains both the
vector and axial-vector response functions and is given by
\begin{equation} 
 R_{\alpha\beta}(-\omega,|\vec{k}|)= -c_V^2~\Im 
 m[\Pi^V_{\alpha\beta}(\omega,|\vec{k}|)] - c_A^2~~\Im 
m[\Pi^A_{\alpha\beta}(\omega,|\vec{k}|)]\,.
\end{equation}  

In the non-relativistic limit, we shall focus on density fluctuations and ignore velocity fluctuations. In this case the vector-polarization function $\Pi^V_{\alpha\beta}(\omega,|\vec{k}|) = \delta_\alpha^0~ \delta_\beta^0~ \Pi_0(\omega,|\vec{k}|)$ where $\Pi_0(\omega,\vec{k})$ is the density-density correlation function \cite{FetterWalecka} given
by
\begin{equation}
\Pi_0(\omega,|\vec{k}|)=-{i} \int d^4x~e^{-{i} 
(\vec{k} \cdot \vec{x}- \omega t) } 
Tr(\rho_G ~[ \rho(x,t),\rho(0,0)] )\,,
\label{polar_den}
\end{equation}
where $\rho_G$ is the density matrix and $\rho(x,t)$ is the density
operator. Similarly the axial response in the non-relativistic limit is dominated by spin fluctuations and we can write $\Pi^A_{\alpha\beta}(\omega,|\vec{k}|) = \delta_\alpha^i~ \delta_\beta^j~ \Pi_{ij} (\omega,|\vec{k}|)$ where $i,j=1,2,3$ and $\Pi_{ij}(\omega,\vec{k})$ is the spin correlation function \cite{FetterWalecka} given
by
\begin{equation}
\Pi_{ij}(\omega,|\vec{k}|)=-{i} \int d^4x~e^{-{i} 
(\vec{k} \cdot \vec{x}- \omega t) } 
Tr(\rho_G ~[ \sigma_i(x,t),\sigma_j(0,0)] )\,.
\label{polar_spin}
\end{equation}
The diagonal components in Eq.~\ref{polar_spin} are equal and is denoted by $\Pi_{\sigma}$, while the off-diagonal components of $\Pi_{ij}$ do not contribute to the emissivity of an isotropic medium \cite{Yakovlev:1999}.  We can therefore write the neutrino emissivity as 
\begin{equation}
\dot{\epsilon}_{\nu\bar{\nu}}= \frac{G_F^2}{192~\pi^5} 
\int\! d^3 \vec{k}~k^2~ \left[ c_V^2~I_{\rho}(k) + 3 c_A^2~I_{\sigma}(k) \right]  \,,
\end{equation}
where
\begin{eqnarray}
I_{\rho}(k) &=&    -\int_{k}^{\infty}d\omega\quad  \frac{\omega}{\exp{\left(\beta \omega\right)} -1} ~\Im m[\Pi_0(\omega,|\vec{k}|)] \,,\\
I_{\sigma}(k) &=& -\int_{k}^{\infty}d\omega\quad  \frac{\omega}{\exp{\left(\beta \omega\right)} -1}~\left(\frac{\omega^2}{k^2}-\frac{2}{3}\right)~\Im m[\Pi_\sigma(\omega,|\vec{k}|)] \,.   
\end{eqnarray} 

%The emissivity due to density fluctuations is then given by

%\begin{equation}
%\dot{\epsilon}_{\nu\bar{\nu}}= \frac{-G_F^2~c_V^2}{192~\pi^5} 
%\int\! d^3 \vec{k}\int_k^{\infty}d \omega ~ \frac{\omega~~k^2 ~}
%{(e^{\beta \omega} -1)}~\Im m~\Pi_0(\omega,k) \,.
%\end{equation} 
\section{Vector Response} 
First, we calculate the vector-current response function to verify and understand the nature of suppression factors found in Refs.~\cite{LeinsonPerez:2006a,Sedrakian:2006}. Calculations of the superfluid density response has a long history in condensed matter physics and a pedagogic discussion can be found in Ref.~\cite{Schrieffer:1964}.  Here we describe neutron matter at low density with the model Hamiltonian
\begin{equation} 
{\cal H}= \sum_{p,{\rm spin}}~ \xi_p a_{k \uparrow}^\dagger 
a_{k \uparrow} + V \sum_{p,p'} a_{k \uparrow}^\dagger 
a_{-k \downarrow}^\dagger   a_{k \uparrow} a_{-k \downarrow} \,, 
\label{eqn:Hamiltonian}
\end{equation}
where $V$ is the effective four-fermion interaction. 
We regulate the short-range interaction by using a momentum cut-off. The re-normalization scheme is implemented by specifying the gap and
requiring that $V(\Lambda)$ satisfy the gap equation at zero
temperature
\begin{equation} 
\Delta = - V(\Lambda) \int_0^\Lambda~  \frac{d^3p}{(2 \pi )^3}~
\frac{\Delta}{2 E_p} \,.
\end{equation}  
To describe neutron matter at densities of relevance to the crust, we
choose to display results at $k_F^2/(2m)=\mu=30$ MeV and a momentum
cut-off of $\Lambda=2 k_F$. Though we are ultimately interested in
calculating the PBF rate in the temperature range $T\sim 1-10 \times
10^8$ K and near $k_F \lsim 100$ MeV where the gap $\Delta \lsim 0.1$
MeV \cite{Hebeler:2006kz}, our numerical results are primarily at
slightly larger densities and for $\Delta=1$ MeV where the numerical
computations are somewhat easier. Although finite-range effects
of the nucleon-nucleon interaction are relevant in computing the magnitude of the pairing gap in the neutron star crust here we will restrict our analysis to a simple zero-range interaction but a strength adjusted to reproduce the pairing gap in more sophisticated calculations \cite{Gezerlis:2007}. 

We define the polarization tensor in the mean field approximation by 
\begin{eqnarray}
\Pi^V_{\alpha\beta}(\omega,|\vec{k}|)=  -i\int \frac{d^4p}{(2 \pi )^4} 
{\rm Tr}\left[ \gamma_\alpha G(p+k) \gamma _\beta  G(p)  \right] \,, 
\label{eqn:meanfield1}
\end{eqnarray} 
where $\gamma_\alpha=\left(\tau_3, \hat{1}~(\vec{p}+\vec{k}/2)/M
\right )$. Here the zero-zero component of $\Pi^V_{\alpha,\beta}$ corresponds to the density-density response function defined in Eq.~\ref{polar_den}. Explicitly this is given by 
\begin{eqnarray}
\Pi_{\rm MF}(\omega,|\vec{k}|)=  -i\int \frac{d^4p}{(2 \pi )^4} 
{\rm Tr}\left[ \tau_3 G(p+k) \tau_3  G(p)  \right] \,,
\label{eqn:meanfield2}
\end{eqnarray} 
where the quasi-particle propagator in the $^1S_0$ superfluid state is
given by
\begin{equation}
G(p)=\frac{p_0~\hat{1} + \xi_p ~\tau _3 + \Delta ~\tau _1}
{p_0^2-E_p^2+i \epsilon}\,,
\end{equation}
and $\hat{1}$ is the unit matrix, $\tau_{i=1,2,3}$ are the $2 \times
2$ Pauli matrices (acting in the Nambu-Gorkov space). The
quasi-particle energy is $E_p=\sqrt{\xi_p^2+\Delta^2}$ where
$\xi_p=(p^2/2m - \mu)$ and $\Delta$ is the superfluid gap
\cite{Schrieffer:1964}. 

\begin{figure}[ht]
\includegraphics[width=\columnwidth]{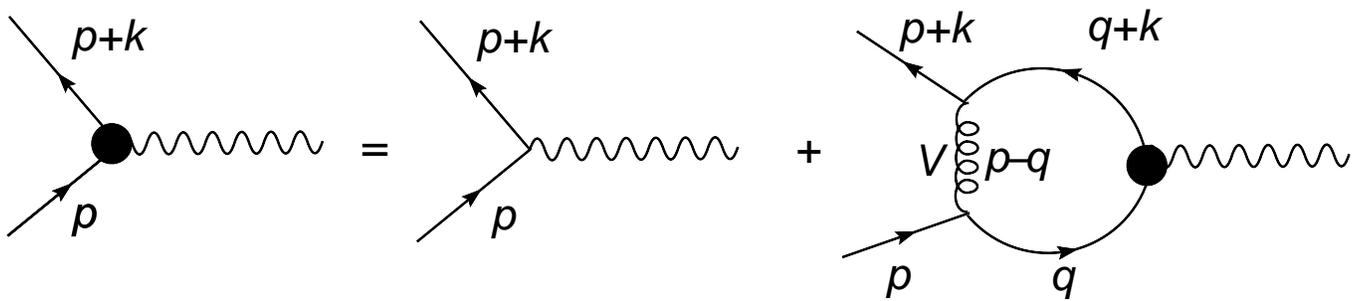}
\caption{Diagrammatic representation of the RPA vertex equation. 
The wavy line is the external weak current and curly line represents the short-range strong interaction which in this work is simplified to a point interaction. Solid lines 
represent the quasi-particles and the dark circle is the dressed vertex.}
\label{vertex}
\end{figure}

\begin{figure}[ht]
\includegraphics[width=0.4\columnwidth]{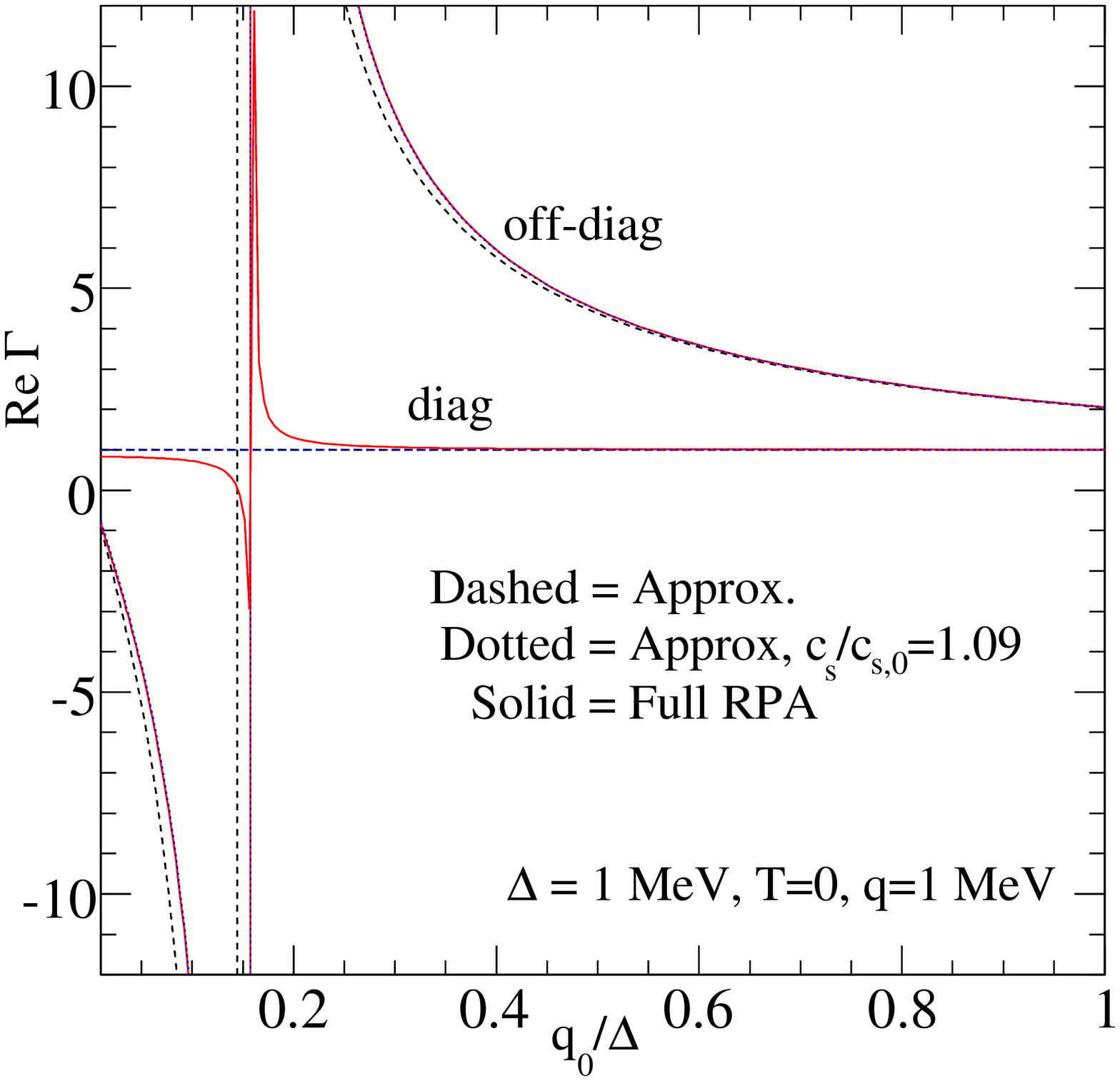}
\includegraphics[width=0.4\columnwidth]{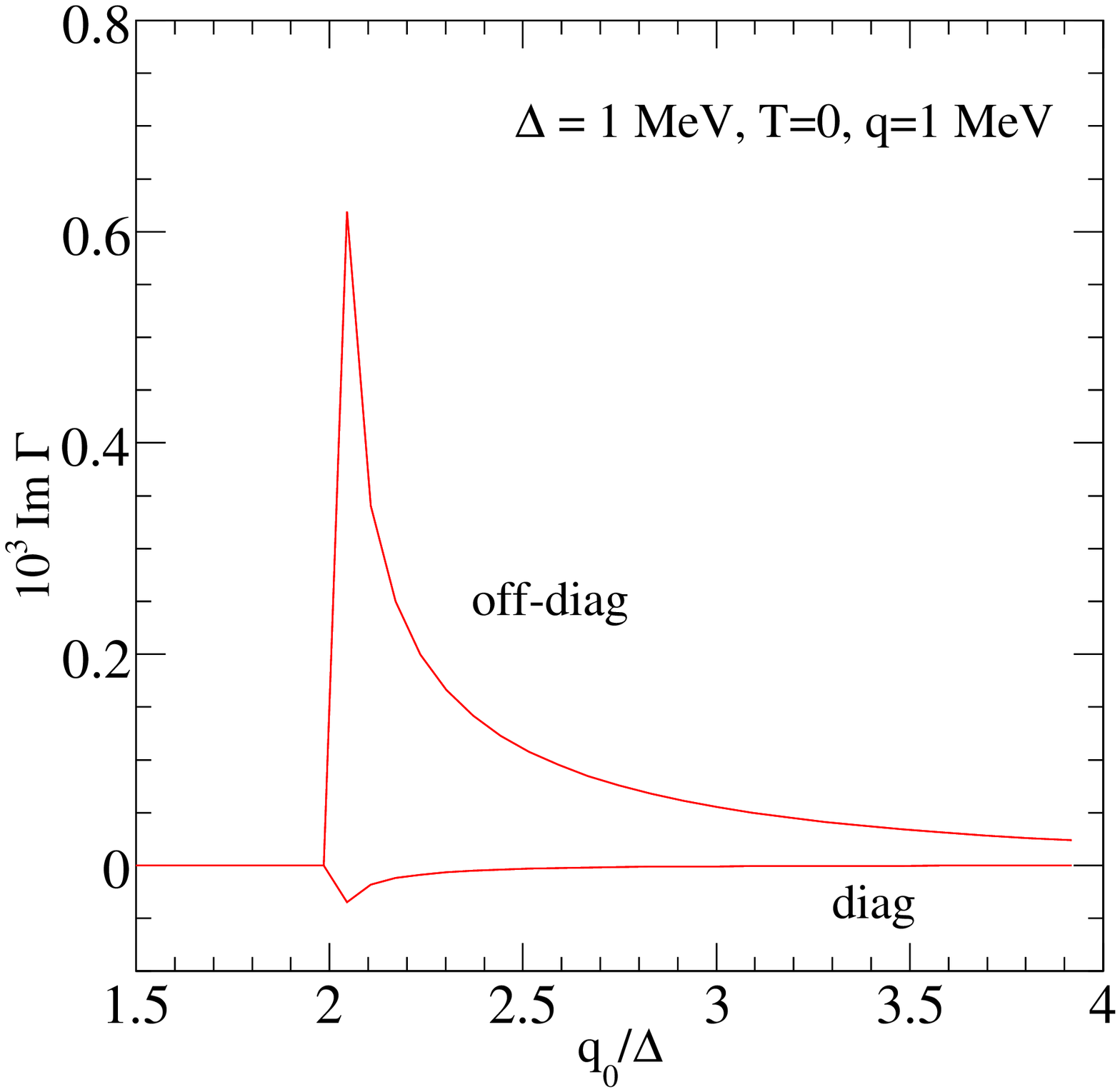}
\caption{Real parts of the
dressed vertex function at fixed momentum transfer. The approximate solution in Eq.~\ref{eqn:approx_kappa} is in excellent agreement with the "exact" RPA vertex especially when the 
speed of sound is shifted to match the pole structure. Right panel shows the imaginary part of the vertex function is finite but small. The approximate vertex  assumes that imaginary part is zero. }
\label{fig:dressedvertex}
\end{figure}

The mean-field polarization tensor violates current conservation and the F-sum rule. This is a well established finding and a lucid discussion can be found in the original papers by Anderson \cite{Anderson:1958} and Nambu \cite{Nambu:1960}. To restore current-conservation it is necessary to replace the bare vertex function $\gamma_{\alpha}(p+q,p)$ by the dressed vertex $\Gamma_{\mu}(p+k,p)$ corresponding the dressed quasi-particles in the superfluid. The dressed vertex is then required to satisfy the generalized Ward-identity (GWI) given by
\begin{equation} 
\omega \Gamma^0 (p+k,p) - k_i\Gamma^{i}(p+k,p) = G^{-1}(p+k)\tau_3 - 
\tau_3 G^{-1}(p)  \,,
\label{eqn:gwi}
\end{equation}  
where $i=1,2,3$ and as before the four-vector $k=(\omega, \vec{k})$.
However the GWI does not uniquely determine the vertex function. It
must be obtained explicitly as a solution to an integral
equation which describes the modification of the vertex
from the medium. This integral equation for the vertex, obtained
in the random-phase approximation (RPA), is known to satisfy the GWI
and is diagrammatically in Fig.\ref{vertex}. Explicitly this is given
by
\begin{equation} 
\Gamma_\alpha = \gamma_\alpha + i~V~\int~\frac{d^4q}{(2\pi)^4}~
\tau_3~G(q+k)~\Gamma_\alpha~G(q)~\tau_3 \,.
\label{eqn:vertex}
\end{equation} 
%==============================================
%===========================================
We are interested in the zeroth-component of Eq.\ref{eqn:vertex} since it is this that affect the density response. In this case an approximate solution has the form
\begin{equation}
\Gamma_0=  \frac{1}{1+\chi}~ \tau_3 ~+ ~\frac{2~\kappa}{1+\chi}~{i}~\tau_2
\label{eqn:approx_vertex}
\end{equation}
In weak coupling, $\chi \simeq V~N(0) \ll 1$ where $V$ is the
four-fermion coupling and $N(0) \propto M~k_F/\pi$ is the density of
states at the Fermi surface and can be neglected. However $\kappa$ has
an essential singularity at $T=0$ corresponding to existence of a
Goldstone excitation that couples to density fluctuations. At $T=0$ and weak coupling   
\begin{equation} 
\kappa \simeq   \frac{\Delta~\omega}{\omega^2-c_s^2~k^2} \,,
\label{eqn:approx_kappa}
\end{equation} 
where speed of the Goldstone mode $c_s \simeq k_F/\sqrt{3}M$. This approximate form for the vertex 
was used in Ref.~\cite{LeinsonPerez:2006b} to compute the response function. It is possible to solve the vertex equation (Eq.~\ref{eqn:vertex}) to obtain the exact solution in the random phase approximation \cite{Kundu:2004mz}.  In Fig.~\ref{fig:dressedvertex} we compare the exact solution (solid-line) to the vertex equation to the approximate expression above (dashed-line). For a better approximation to the full vertex, we can employ a shifted value of $c_s$ by shifting it so that the
approximate and full vertices diverge at exactly the same value of
$q_0$. This ``shifted'' vertex is also plotted in
Fig.~\ref{fig:dressedvertex} with a dotted line, but the result
matches sufficiently well with the full vertex that it is not visible.
We also find that this relative agreement between the real parts of the full and approximate
results is not strongly modified at finite temperature.
%==============================================
%===========================================
In the region where $\omega \ge 2 \Delta$ where we are above the threshold for producing quasi-particles we can expect that the vertex equation will have a non-zero imaginary part. The approximate solution to the vertex equation in Eq.~\ref{eqn:approx_kappa} neglects this contribution. The imaginary part of the full vertex is plotted in the right panel of Fig.~\ref{fig:dressedvertex}. As expected below $\omega=2 \Delta$, the imaginary part of the response vanishes as required from the delta function given by the imaginary part of the propagator in Eq.~\ref{eqn:mfresponset0}. The magnitude of the imaginary part is much smaller than the real part, but as we shall discuss later could make an be  important contribution to the response.

To help make contact with earlier results obtained in Ref.~\cite{FRS:1976,LeinsonPerez:2006a} we discuss the response 
function in different approximations. First, we obtain the mean field response at $T=0$ by doing the $p_0$ integration in
Eq.~\ref{eqn:meanfield2}
\begin{eqnarray}
\Pi_{\rm MF}(\omega,|\vec{k}|) = 
\int \frac{d^3p}{2 (2\pi)^3}~  \left(1-\frac{\xi_p \xi_{p+k}-\Delta
^2} {E_p E_{p+k}}\right)~I_0 &&\nonumber \\
I_0= \left(\frac{1}{\omega-E_p-E_{p+k}+i \epsilon} - \frac{1}
{\omega+ E_p+E_{p+k}-i\epsilon }\right) &&
\label{eqn:mfresponset0}
\end{eqnarray}
In the long-wavelength limit ($k \rightarrow 0$) this can be simplified further and 
we find that 
 \begin{equation}
\Pi_{\rm MF}(\omega,|\vec{k}|\rightarrow 0) =  \int
\frac{d^3p}{(2\pi)^3} 
~\frac{\Delta^2}{E_p^2}~I_0 \,.
\end{equation}
For $\omega> 2 \Delta$ and $k=0$ the mean field result predicts a non-vanishing response given by  
\begin{equation}
\Im m[\Pi_{\rm MF}(\omega)] = -\frac{Mp_F}{\pi^2}~\left( 
\frac{\Delta^2}{\omega~\sqrt{\omega^2-4 \Delta^2}}~+\frac{\Delta^2}
{4\mu~\omega}\right) \,.
\end{equation}
As mentioned earlier a non-zero response at $k=0$ violates current
conservation and the related F-sum rule given by 
\begin{equation}
\int_{-\infty}^{\infty}~d \omega ~ \omega~\Im m[\Pi_{\rm MF}(\omega)] = \langle [[H,\rho(k)],\rho(k)]\rangle \,,
\label{eqn:vfsum}
\end{equation} 
where the RHS vanishes in the $k=0$ limit when $\rho$ commutes with the Hamiltonian. This is simply a consequence of the well known fact that radiation can arise only as a result of particle acceleration for conserved charges. The vertex correction discussed earlier remedies this problem. The RPA response function is defined by 
\begin{equation}
\Pi_{\rm RPA}(\omega,|\vec{k}|)=  -i\int \frac{d^4p}{ (2 \pi )^4} 
{\rm Tr}\left[ \tau_3 G(p+k) \Gamma_0 G(p)  \right] \,,
\label{eqn:RPA}
\end{equation} 
where $\Gamma_0$ is the dressed vertex that satisfies Eq.~\ref{eqn:vertex}. First we use the approximate form of the vertex given in Eq.~\ref{eqn:approx_kappa} to obtain the zero-temperature density-density polarization function. As we showed earlier the approximate vertex 
\begin{equation} 
\Gamma_0=\tau_3 + \frac{2 \Delta \omega}{\omega^2 - c_s^2 k^2}~{i} \tau_2
\end{equation} 
provides a very good description of the real part of the exact result. Substituting this into Eq.~\ref{eqn:RPA} and performing the integration over the $p_0$ we obtain
\begin{equation}
\Pi_{\rm RPA}(\omega,k)=
\int \frac{d^3p}{2 (2\pi)^3} ~ \left(1-\frac{\xi_p \xi_{p+k}-\Delta ^2+2 
\omega \kappa\Delta}{E_p E_{p+k}}\right)~I_0 \,. 
\end{equation} 
The response function is related to the imaginary part which is explicitly given by 
\begin{equation}
\Im m[\Pi_{\rm RPA}(\omega,k)]= -\int \frac{d^3p}{2 (2\pi)^3}~ J~\delta(\omega-E_p-E_{p+k}) \quad {\rm where}\quad 
J= \left(1-\frac{\xi_p \xi_{p+k}-\Delta ^2+2 
\omega \kappa\Delta}{E_p E_{p+k}}\right)
\label{eqn:rparesponset0}
\end{equation} 
It is straightforward to verify that the imaginary part of $\Pi_{\rm RPA}$ vanishes in the limit
$k\rightarrow 0$ as required by current conservation. At finite temperature and for the full vertex this continues to hold and the response vanishes at $k=0$ because of the gap equation. The critical question then is to inquire if the order $k^2$ term also vanishes. To address this we expand the zero-temperature RPA response in Eq.~\ref{eqn:rparesponset0} in powers of $k$.  Expanding the integrand in 
Eq.~\ref{eqn:rparesponset0} we obtain 
\begin{equation} 
J= \frac{c_s^2~\Delta^2}{2 E_p^4}~\left[ \frac{p^2~x^2}{m^2~c_s^2}-1\right] ~k^2 ~+~ {\mathcal O} \left[ c_s~k\right]^4
\label{eqn:jofk}
\end{equation}
where $x$ is the cosine of the angle between the momenta $p$ and $q$. In the long-wavelength 
most of the support from the energy delta function is near $p=p_F$. Further, in this region we can approximately replace $x^2$ by its mean value given $ x^2 \simeq <x^2> = 1/3$. This implies that  
the quadratic term nearly vanishes since $c_s \equiv p_F/(m\sqrt{3})$. The nature of this cancellation 
depends on the value of $c_s$ employed in the approximated vertex. When we use the self-consistent RPA vertex obtained by solving Eq.~\ref{eqn:vertex} we indeed find that this cancellation is nearly exact and 
Eq.~\ref{eqn:jofk} receives contributions at order $c_s^4~k^4$. 

\begin{figure}[ht]
\includegraphics[width=0.7\columnwidth]{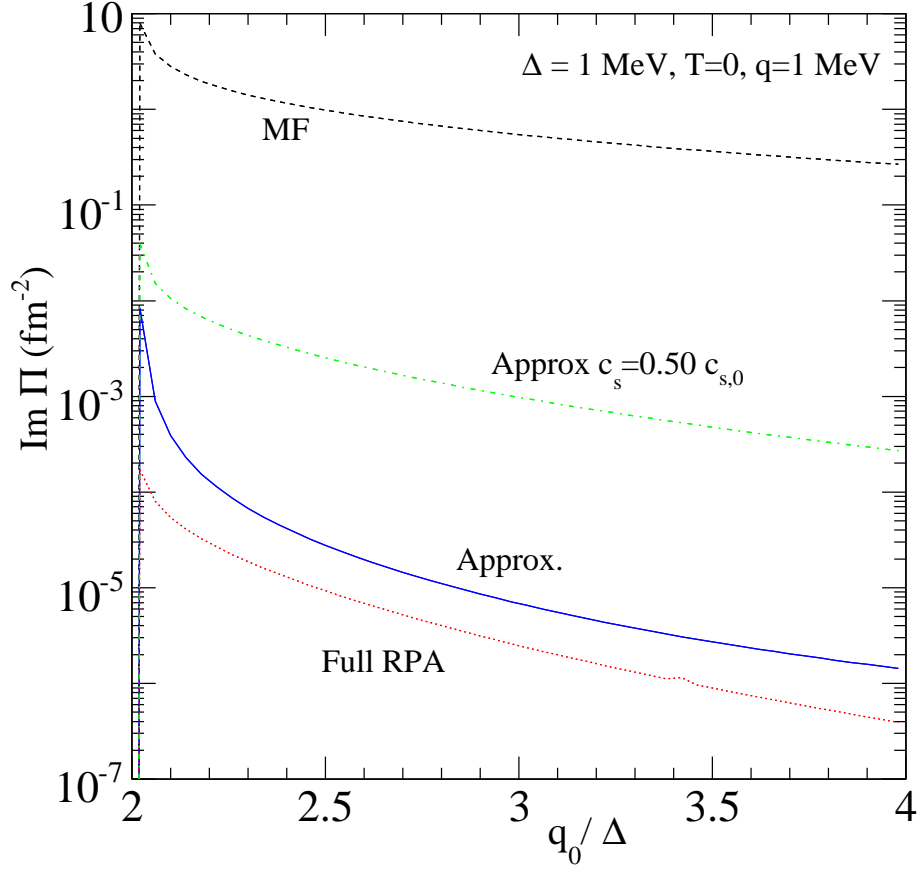}
\caption{Imaginary part of the density-density polarization at $T=0$ obtained using different approximations.  The vertex corrections induce a large suppression in the response relative to the mean-field result. The result obtained using the self-consistent RPA vertex (shown with the dotted line) is suppressed by a factor $\sim 10^{-4}$. See text for a discussion relating to the response obtained using the approximate vertex.}
\label{fig:imrespt0}
\end{figure}

The imaginary part of the response function at zero temperature obtained in different approximate schemes are shown in Fig.~\ref{fig:imrespt0}. The striking feature is that results obtained using the dressed vertex are suppressed because they are $\propto c_s^2$ and the response using the self-consistent RPA vertex labeled "Full RPA" in the figure is even more suppressed because the relevant contribution occurs at order $c_s^4$.  The result labeled "Approx." is obtained using $c_s=c_{s,0}=p_F/(\sqrt{3} M)$. We note that the density response function calculated within RPA coincides with earlier calculations by Kundu and Reddy Ref.~\cite{Kundu:2004mz} \footnote{ Although our results for the response function and RPA vertex derived are essentially the same as in Ref.~\cite{Kundu:2004mz} our numerical results differ because of a numerical error in Ref.~\cite{Kundu:2004mz}.}.    
%==============================================
%===========================================
From the Fig.~\ref{fig:imrespt0} it is clear that the magnitude of the suppression obtained using the approximate vertex is sensitive to $c_s$. This is because of the large cancellation at order $c_s^2~k^2$. To understand the nature of this cancellation we first note that vector current conservation does not require the order  $v_F^2~k^2$ contribution to vanish. This would require fine tuning the velocity of the Goldstone mode to precisely cancel the contribution at order $k^2$. In general, we are aware of no mechanism that can accomplish this in a multicomponent system. 

The preceding arguments raises the following question: {\it what is the underlying physics responsible for the large cancellation at order $k^2$ occurring in the RPA calculation of the response function for pure neutron matter ?}. To address this question we note that a similar cancellation at order $k^2$ occurs in neutrino emission due to density fluctuations in the bremsstrahlung reaction $nn \rightarrow nn \nu \bar{\nu}$ in the normal phase. Here the square of the matrix element contributes only at order $v_F^4 k^4$ \cite{FrimanMaxwell:1979,vanDalen:2003}. This result is well understood, especially in the context of electromagnetic bremsstrahlung in proton-proton collisions where radiation occurs due to time variation of the quadrupole  moment. The dipole radiation vanishes because the dipole moment for {\it identical} particles interacting with each other does not vary in time. In fact for any system of N identical particles with charge ${e}$ and mass m, the dipole moment
\begin{equation}
\vec{d} = \Sigma _i {e} \vec{r}_i = \frac{{e}}{m}
\Sigma _i m~\vec{ r_i} =\frac{{e}}{m}\Sigma _i m~\vec{R}_{\text{CM}} 
\end{equation}
does not vary because momentum conservation ensures that
$\dot{R}_{\text{CM}}=0$. In the quantum mechanical calculation of the bremsstrahlung process this arises due to the 
destructive interference between the amplitudes for radiation from the two charges. This interference ensures that the  
square of the matrix element vanishes at order $v_F^2 ~k^2$  and the leading contribution is quadrupolar and occurs at order $v_F^4~k^4$. 

When the (weak) charge to mass ratio of the two particles is different, the cancellation at order $v_F^2~k^2$ would be absent. For example bremsstrahlung from neutron-proton scattering occurs at order  $v_F^2~k^2$. Hence we argue in multicomponent systems there is {\it no} symmetry that requires the sensitive cancellation at this order.   In the neutron star context, where neutrons in the crust couple to a background lattice of neutron-rich ions, we can expect that these interactions can induce a response at order $c_s^2~k^2$.  In our calculation, such interactions can for example induce a shift in the superfluid velocity $c_s$  by the polarization of the lattice. As we have seen in Fig.~\ref{fig:imrespt0}, even a small shift leads to a relevant contribution at order $c_s^2k^2$.  Thus we conclude that in realistic situations a non-vanishing density response at order $c_s^2 k^2$ is to expected. However, as we show below this response is still 
small compared to the axial-current response.    
  
\section{Axial Current Response}  
\begin{figure}[ht]
\includegraphics[width=0.7\columnwidth]{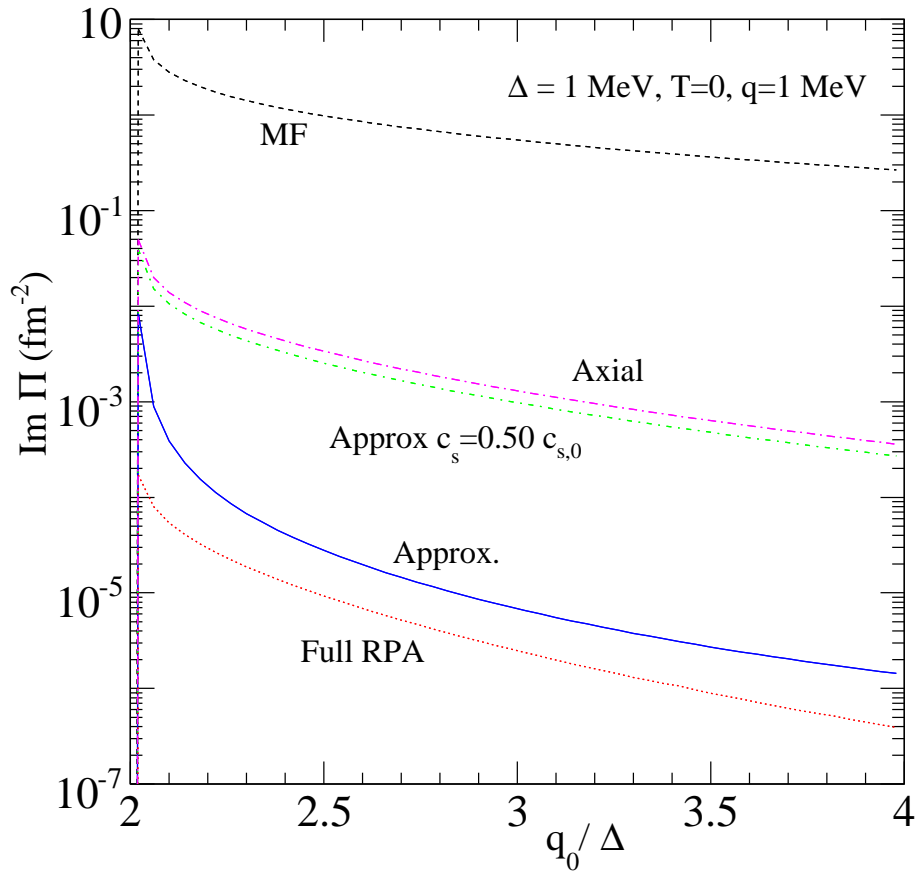}
\caption{The imaginary part of the axial response function in the non-relativistic limit. }
\label{fig:axial}
\end{figure}

In the non-relativistic limit, the diagonal part of the axial polarization tensor defined in Eq.~\ref{polar_spin} can be written in terms of the Nambu-Gorkov propagators \cite{Kundu:2004mz}. To begin we will ignore vertex corrections because there is no Goldstone mode associated with spin-fluctuations in the case of singlet pairing and discuss the one-loop mean field polarization tensor.  In this case we can write 
\begin{eqnarray}
\Pi_{\sigma}(\omega,|\vec{k}|)=  -i\int \frac{d^4p}{(2 \pi )^4}~ 
{\rm Tr}\left[ \hat{1} ~G(p+k) ~\hat{1}~  G(p)  \right] \,. 
\label{eqn:axial}
\end{eqnarray} 
At $T=0$ we can do the $p_0$ integration to obtain 
\begin{equation}
\Pi_{\sigma }(\omega,|\vec{k}|) = 
\int \frac{d^3p}{2 (2\pi)^3}~  \left(1-\frac{\xi_p \xi_{p+k}+\Delta
^2} {E_p E_{p+k}}\right)~I_0 \,,
\label{eqn:axialt0}
\end{equation}
and the imaginary part in the region $\omega \ge 0$ is given by
\begin{equation}
\Im m[\Pi_{\sigma }(\omega,|\vec{k}|)] = 
-\int \frac{d^3p}{2 (2\pi)^3}~  \left(1-\frac{\xi_p \xi_{p+k}+\Delta
^2} {E_p E_{p+k}}\right)~\delta(\omega - E_p - E_{p+k}) \,.
\label{eqn:imaxialt0}
\end{equation}
It is easily verified that the imaginary part of Eq.~\ref{eqn:imaxialt0} vanishes at $k=0$.  Expanding the integrand in Eq.~\ref{eqn:imaxialt0} in powers of $k$ we find that  
\begin{equation}
\Im m[\Pi_{\sigma }(\omega,|\vec{k}|)] = -\frac{1}{32\pi^2}
\int dp~p^2\int dx ~\left(\frac{p^2}{m^2}~\frac{\Delta^2}{E_p^4} ~x^2~k^2+ {\mathcal O}[k^4]\right) 
~\delta(\omega - E_p - E_{p+k}) \,, 
\label{eqn:newlabel}
\end{equation}
where $x$ is the angle between $\vec{k}$ and $\vec{p}$.  In the long-wavelength limit the delta function provides support only in region $p\simeq p_F$.  Here we can further simply the result to obtain
\begin{equation}
\Im m[\Pi_{\sigma }(\omega,|\vec{k}|)] = -\frac{1}{48\pi^2}~v_F^2~k^2~\int dp~p^2~\frac{\Delta^2}{E_p^4}~\int dx  ~\delta(\omega - E_p - E_{p+k})~+ {\mathcal O}[k]^4 \,.
\label{eqn:imaxial2}
\end{equation}
In Fig.~\ref{fig:axial} we plot the axial response function. The vector response obtained in different approximations discussed earlier is also shown for comparison.The results indicate the axial response is significantly larger because of the large cancellation in the vector response at order $k^2$. In situations where this cancellation does not occur, as in the case when we set $c_s \simeq 0.5 c_{s,0}$, the vector and axial response functions can become comparable. 

%==============================================
%===========================================

The one-loop axial response in Eq.~\ref{eqn:axial} vanishes in the $k=0$ limit. This is consistent with the F-sum rule for the spin response associated the model Hamiltonian in  Eq.~\ref{eqn:Hamiltonian}. However it is well-known that the realistic nuclear Hamiltonian which contains both tensor and spin-orbit interactions does not commute with the spin operator. This implies that for realistic nuclear interactions which include pion exchange
\begin{equation}
\int_{-\infty}^{\infty}~d \omega ~ \omega~{\rm lt}_{k \rightarrow 0} 
\Im m[\Pi_{\rm \sigma}(\omega)] = \langle [[H,\sigma(k=0)],\sigma(k=0)]\rangle ~\neq 0 \,.
\label{eqn:afsum}
\end{equation} 
Consequently, the correct axial response obtained using realistic interactions has to be finite in the limit $k=0$. From Eq.~\ref{eqn:imaxial2} we see that the mean field response function violates this expectation. Corrections to the one-loop axial response function arising due to non-central interactions such as pion exchange are therefore will therefore be critical in the long-wavelength limit.

The importance of pion exchange in the axial response and neutrino emissivity was already realized in the pioneering work of Friman and Maxwell~\cite{FrimanMaxwell:1979}. Although their calculations were only applicable to the normal phase of neutron matter, they showed that pion exchange was more important than central nuclear interactions. In particular, they found that the matrix element for neutrino emission from neutron-neutron bremsstrahlung in the axial channel was non-zero in the long-wavelength limit when they included one-pion exchange. It is therefore critical to incorporate these non-central interactions in the calculation of the axial response function of the neutron superfluid. This would require that we include both vertex corrections and two-loop effects which include $2p-2h$ excitations in the superfluid. This work is in progress and will be reported elsewhere.  

\section{Neutrino Emissivity}

\begin{figure}[ht]
\includegraphics[width=0.5\columnwidth]{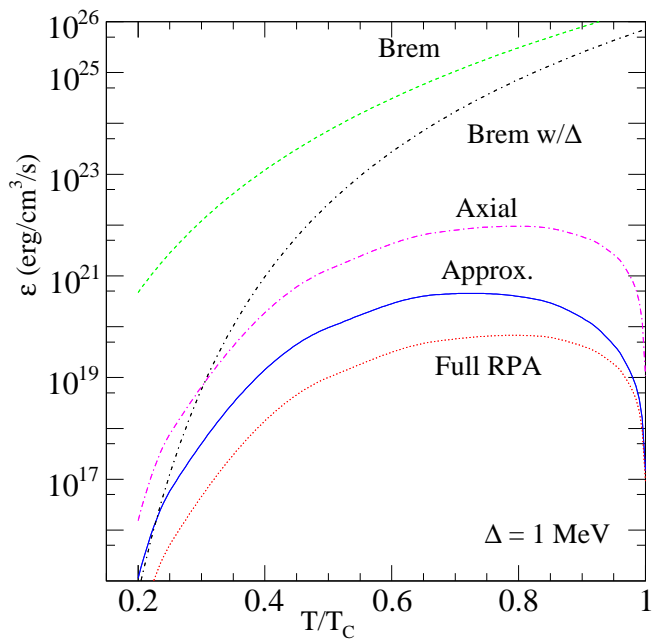}
\caption{The neutrino bremsstrahlung and pair-breaking emissivities.
The dashed line is the bremsstrahlung emissivity in non-superfluid
neutron matter and the short dashed-dotted line is the corresponding
emissivity in superfluid neutron matter. The long dashed-dotted line
is the axial part of the pair-breaking emissivity and the dotted 
line is the vector part of the pair-breaking emissivity using the 
full RPA vertex. The solid line is the vector pair-breaking emissivity 
using the approximate vertex.}
\label{fig:em}
\end{figure}

The neutrino emissivity for the processes discussed above is plotted in Fig~\ref{fig:em}, including the axial and vector contributions
to the PBF emissivity and the bremsstrahlung emissivity with and without the suppression factors resulting from superfluidity as
obtained in Ref.~\cite{Yakovlev:1999}. The emissivity from the axial response dominates and is two or three order of magnitudes larger than the full RPA vector response. A shift of the speed in sound of the Cooper pair excitations will modify the emissivity quadratically in $c_s$ as expected from the imaginary part of the vector response. The
bremsstrahlung emissivity is larger than the pair-breaking emissivity except at low temperatures when the suppression of the bremsstrahlung from superfluidity is strong. 

%==============================================
%===========================================
 
\section{Conclusions:} 
We have studied the one-loop vector and axial current response functions of relevance to neutrino emission in superfluid neutron matter. Through explicit calculations we found that there is strong suppression of the vector response when vertex corrections are included. This is in close agreement with the findings of Leinson and Perez \cite{LeinsonPerez:2006a}. For pure neutron matter, the RPA vertex function does indeed show that the supression factor is of order $v_F^4$. However we have shown this suppression arises 
not only because of vector current conservation, but also 
because the radiation in the vector channel for pure neutron matter occurs only due to time variation of the quadrupole moment. In the realistic context where neutrons interact with the background lattice of of neutron-rich ions the suppression in the vector channel is of order $v_F^2$. We showed that even a small shift in the speed of the Goldstone mode due to the lattice can make a relevant contribution to the response at this order. Finally, we showed that both of these emissivities are likely smaller than the neutrino bremsstrahlung emissivity (including the suppression factors from superfluidity) unless the temperature is significantly smaller than the critical temperature.

The axial response function was shown to be numerically more important because because the dominant contribution occurs at order  $v_F^2$.  Although this was the relevant contribution in our model, we demonstrated that tensor interactions arising due to pion exchange would lead to important corrections to this estimate. The F-sum rule in this case indicates that these correction would result in a non-vanishing response at order $v_F^0$. In the normal phase the tensor interaction is responsible for the emissivity at long-wavelength and we suspect that this will continue to be the case in superfluid matter. 
We anticipate that tensor interactions will also affect the PBF fluctuations in the spin channel and can modify our results and the regime in temperature where PBF can dominates over the bremsstrahlung emissivity. This issue is currently being investigated and will be reported elsewhere. Although it is now clear that the neutrino emissivity due to density fluctuations arising from PBF processes in the vicinity of the critical temperature is not significant, our present understanding of neutrino processes in the superfluid phase remains rather incomplete and warrants further study. 

{\it Acknowledgments:} The authors would like to that Edward Brown for
useful discussions related to this work. This research was supported
by the Dept. of Energy under contract W-7405-ENG-36, the Joint
Institute for Nuclear Astrophysics at MSU under NSF-PFC grant PHY
02-16783, and by NASA ATFP grant NNX08AG76G.

\vspace{-0.2in}
%=========== Bibliography ==================
\bibliographystyle{h-physrev4}
\bibliography{pbf_bib} 
%==========================================
\end{document}